\documentclass[%
 reprint,
 amsmath,amssymb,
 aps,
]{revtex4-1}

\usepackage{graphicx}
\usepackage{dcolumn}
\usepackage{bm}

\usepackage[titletoc,title]{appendix}  

\newcommand{\ep}{\varepsilon}

\newcommand{\bra}[1]{{\langle #1 |}}
\newcommand{\ket}[1]{{| #1 \rangle}}

\newcommand{\rv}{{\bf r}}

\newcommand{\Rv}{{\bf R}}

\newcommand{\kv}{{\bf k}}
\newcommand{\Kv}{{\bf K}}

\begin{document}

\title{Interlayer Resistance of Misoriented MoS$_2$}

\author{Kuan Zhou}
\email[Email:~]{kzhou003@ucr.edu}
\affiliation{Department of Physics and Astronomy, University of
California, Riverside, CA 92521-0204}
\author{Darshana Wickramaratne}
\affiliation{Materials Department, University of
California, Santa Barbara, CA 93106-5050}
\author{Supeng Ge}
\affiliation{Department of Physics and Astronomy, University of
California, Riverside, CA 92521-0204}
\author{Shanshan Su}
\affiliation{Department of Electrical and Computer Engineering, University of
California, Riverside, CA 92521-0204}
\author{Amrit De}
\affiliation{Department of Physics and Astronomy, University of
California, Riverside, CA 92521-0204}
\author{Roger K. Lake}
\email[Email:~]{rlake@ece.ucr.edu}
\affiliation{Department of Electrical and Computer Engineering, University of
California, Riverside, CA 92521-0204}

\begin{abstract}
Interlayer misorientation in transition metal dichalcogenides alters
the interlayer distance, the electronic bandstructure, and the vibrational modes, 
but, its effect on the interlayer resistance is not known.
This work analyzes the coherent interlayer resistance
of misoriented 2H-MoS$_2$ for low energy electrons and holes as a function
of the misorientation angle.
The electronic interlayer resistance monotonically increases  with the supercell
lattice constant by several orders of magnitude similar to that of misoriented bilayer graphene.
The large hole coupling gives low interlayer hole resistance that weakly depends on the 
misorientation angle. 
Interlayer rotation between an n-type region and a p-type region will suppress 
the electron current with little effect on the hole current.  
We estimate numerical bounds and explain the results in terms of the orbital composition of the bands at high symmetry points.
Density functional theory calculations provide the interlayer coupling used in both a tunneling
Hamiltonian and a non-equilibrium Green function calculation of the resistivity.
\end{abstract}


\maketitle 

{\em Introduction}:
There is tremondous interest in multilayer 
and heterostructure stacks of 
transition metal dichalcogenides (TMDs)
\cite{Geim_Grigorieva_vdWHets_Nat13,
ML_Hets_LKou_JPChem13,
QuGates_BL_TMDs_NatCom13,
E_tune_valley_B_mom_BL_MoS2_Xu_NPhys13,
spin-layer-locking_BL_WSe2_Xu_NatPhys14,
Het_Review_Heinz_Hone_Kim_AIPMat14,
pn_jn_Hone_Heinz_Kim_NatNano14,
Ajayan_WS2_MoS2_NatMat14,
Ajayan_BandEng_TMD_Small14,
VHets_vdW_epi_SJin_NL14,
Hone_Heinz_Tailoring_Twist_MoS2_NL14,
Louie_Zettl_Twist_MoS2_NComm14,
Dresselhaus_twist_MoS2_NL14,
Yablonovitch_Javey_vdW_Hets_PNAS14,
Darshana_JCP14,
Interlayer_phonons_RHe_PRB15,
Heinz_Interlayer_Interactions_NL15,
Cronin_AdvMat15,
Cronin_ACSPhotonics16,
2Step_Growth_WSe2_MoSe2_NL15,
Hinkle_Wallace_MBE_HfSe2_ACSNano15,
Ajayan_Layer_Eng_AdvMat16,
MoSe2_HFSe2_TaSe2_AppMatInt16,
Tongay_Nanoscale_Rev16}.
They exhibit strong spin orbit coupling and non-trivial topology  \cite{QuGates_BL_TMDs_NatCom13,
E_tune_valley_B_mom_BL_MoS2_Xu_NPhys13,
spin-layer-locking_BL_WSe2_Xu_NatPhys14},
large Seebeck coefficients \cite{Darshana_JCP14},
tunable bandstructure \cite{Cronin_AdvMat15,Cronin_ACSPhotonics16,Tunable_TBL_MoS2_Sutter_Hone_Osgood_NL16},
many possibilities for band 
engineering \cite{KCho_Band_Aignments_erratum14},
type II band alignments \cite{Yablonovitch_Javey_vdW_Hets_PNAS14,2Step_Growth_WSe2_MoSe2_NL15,Heinz_Interlayer_Interactions_NL15},
and rectifying pn junctions 
\cite{pn_jn_Hone_Heinz_Kim_NatNano14,Yablonovitch_Javey_vdW_Hets_PNAS14,2Step_Growth_WSe2_MoSe2_NL15}.
Multilayer and heterostructure growth with stacking control  have been demonstrated\cite{2Step_Growth_WSe2_MoSe2_NL15,
Hinkle_Wallace_MBE_HfSe2_ACSNano15,
MoSe2_HFSe2_TaSe2_AppMatInt16,
Ajayan_Layer_Eng_AdvMat16}.
Recent reviews provide an overview of the state of the art
\cite{Xu_Yao_Xiao_Heinz_Valleytronics_Rev14,
Modeling_Stacked_2D_Mats15,
Tongay_Nanoscale_Rev16,
Mak_2D_Photonics_Rev16}.
\begin{figure}
\includegraphics[width=3.5in]{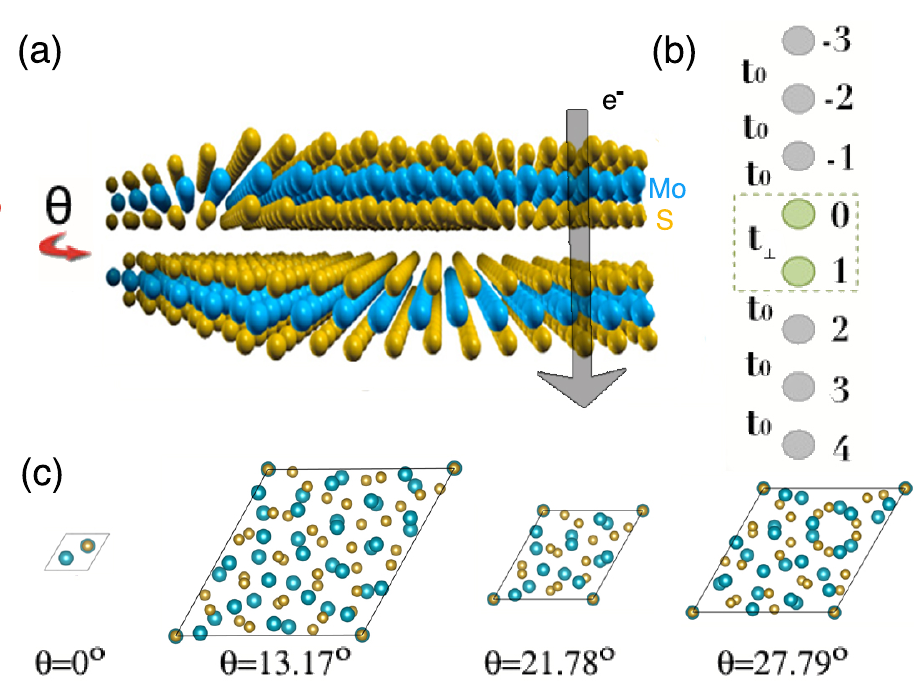}
\caption{
(a) Atomistic geometry of the $21.78^\circ$ rotated interface.
The semi-transparent arrow indicates the direction of current flow.
(b) Reduction to a tight binding chain model for a given
valley and $\kv$.
The two sites 0 and 1 correspond to the two layers.
(c) Commensurate unit cells corresponding to the commensurate
misorientation angles.
\label{fig:TBL_MoS2}
}
\end{figure}

For TMD misoriented bilayers, both experiments and simulations 
show that the interlayer
coupling and the interlayer distance are sensitive to the rotation 
angle, and that the sensitivity of the coupling is very
different for different valleys \cite{Louie_Zettl_Twist_MoS2_NComm14,
Dresselhaus_twist_MoS2_NL14,
Hone_Heinz_Tailoring_Twist_MoS2_NL14,
Interlayer_phonons_RHe_PRB15,
Tunable_TBL_MoS2_Sutter_Hone_Osgood_NL16}. 
A small rotation angle in hetero-bilayers alters the inter-layer exciton 
dynamics \cite{X_Xu_InterLayer_excitons_PRL15,X_Xu_InterLayer_excitons_Sci16}.
While the effects of misorientation on the geometry, electronic bandstructure, and vibrational modes 
of bilayer TMDs have received significant attention, 
the effect of misorientation on the interlayer resistivity of TMDs has not yet
been studied. 
Recent work considered the effect of misorientation on the in-plane transport \cite{AWGhosh_MoS2_twisted_arxiv16}.
In this work, we theoretically determine the coherent electron and hole 
interlayer (vertical) conductance of a misoriented MoS$_2$ interface
as illustrated in Fig. \ref{fig:TBL_MoS2}(a).
In MoS$_2$ bilayers, the low-energy electron transport takes place at the 
$K$ valley, 
and the low-energy hole transport takes place at the $\Gamma$ valley. 
This results in an extremely asymmetric response of the electron
and hole interlayer conductivity to the interlayer misorientation angle.
The coherent interlayer electron transport is exponentially suppressed 
by the misorientation,
and the hole transport is only slightly affected.

{\em Theoretical Methods}:
The structures considered are a 2H aligned bilayer and misoriented bilayers 
with commensurate rotation angles.
The commensurate unit cells are shown in Fig. \ref{fig:TBL_MoS2}(c),
and they are constructed following the method described in 
Ref. [\onlinecite{shallcross2010electronic}].
The rotation angles are 
$\theta ={13.17}^{\circ}$, ${21.78}^{\circ}$, and ${27.79}^{\circ}$, and
the corresponding unit cell lattice constants are 
$\sqrt { 19 } { a }_{ 0 }$, $\sqrt { 7 } { a }_{ 0 }$, $\sqrt { 13 } { a }_{ 0 }$
where $a_0$ is the lattice constant of 2H aligned bilayer equal to 3.16 \AA \cite{Van_de_Walle_MoS2_PRB14} .
Details of the DFT simulation input parameters and output results using the 
Vienna Ab initio Simulation Package (VASP) \cite{VASP:I:PRB:1993,VASP2,VASP:II:PRB:1996,perdew:1996:PBE,Ernzerhof_JChem99} are provided in the Appendix.

The purpose of these DFT calculations is to determine
the energy splitting of the 
band edges resulting from the interlayer coupling.
The DFT calculations are intentionally performed in the absence of
spin-orbit coupling (SOC) to cleanly extract the band splitting from the
interlayer coupling \cite{Cronin_ACSPhotonics16}.
In the absence of SOC, the 
energy splitting $\Delta_\nu(\kv)$
of each band $\nu$ at wavevector $\kv$
due to the interlayer coupling $t_{\perp}^\nu (\kv)$ 
is $\Delta_\nu(\kv) = 2 |t_{\perp}^\nu (\kv)|$.
In the basis of the eigenstates of the individual monolayers,
the low-energy bilayer Hamiltonian 
for each band $\nu$ is
\begin{equation}
H=\begin{pmatrix} 
\varepsilon_\nu(\kv)  & {t}^\nu_{\perp}(\kv) \\ 
{t}^\nu_{\perp}(\kv) & \varepsilon_\nu(\kv)  
\end{pmatrix}
\label{eq:bilayer_HD}
\end{equation}
where $\varepsilon_\nu(\kv)$ is the low-energy
two-dimensional dispersion of band $\nu$.
%

%

\begin{figure}
\includegraphics[width=3in]{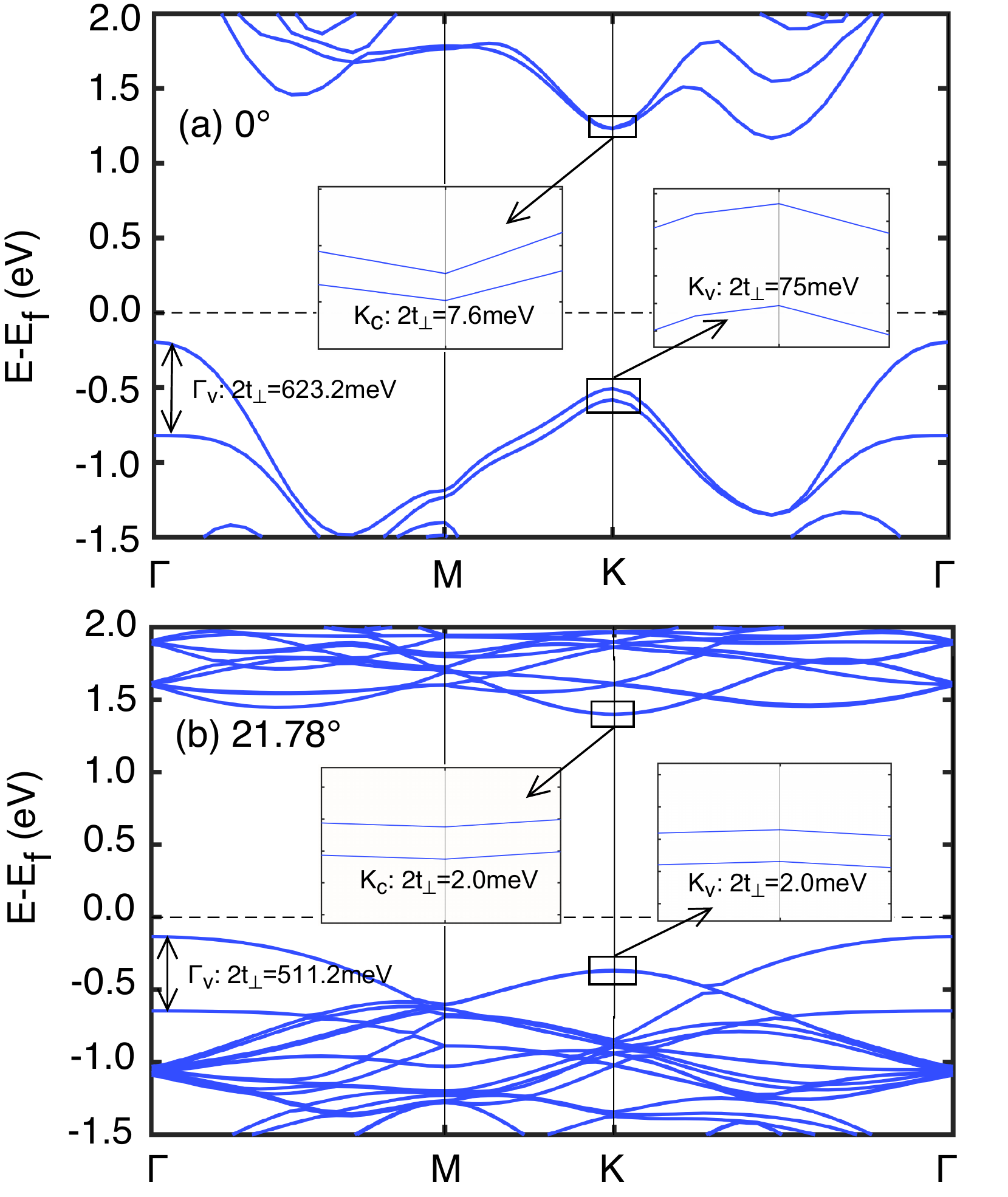}
\caption{
DFT calculated band structures of  (a) unrotated bilayer and (b) ${21.78}^{ \circ  }$ 
misoriented bilayer MoS$_2$ in the absence of SOC.
The splittings of the bands due to interlayer coupling are shown in the insets at the $K$ points,
and directly on the plots at the $\Gamma$ points.
The interlayer coupling parameters $t^\nu_\perp(\kv)$ are extracted from the DFT calculations of the bilayer
electronic bandstructures in the absence of spin orbit coupling.
\label{fig:bands}
}
\end{figure}

\begin{figure}
\includegraphics[width=3in]{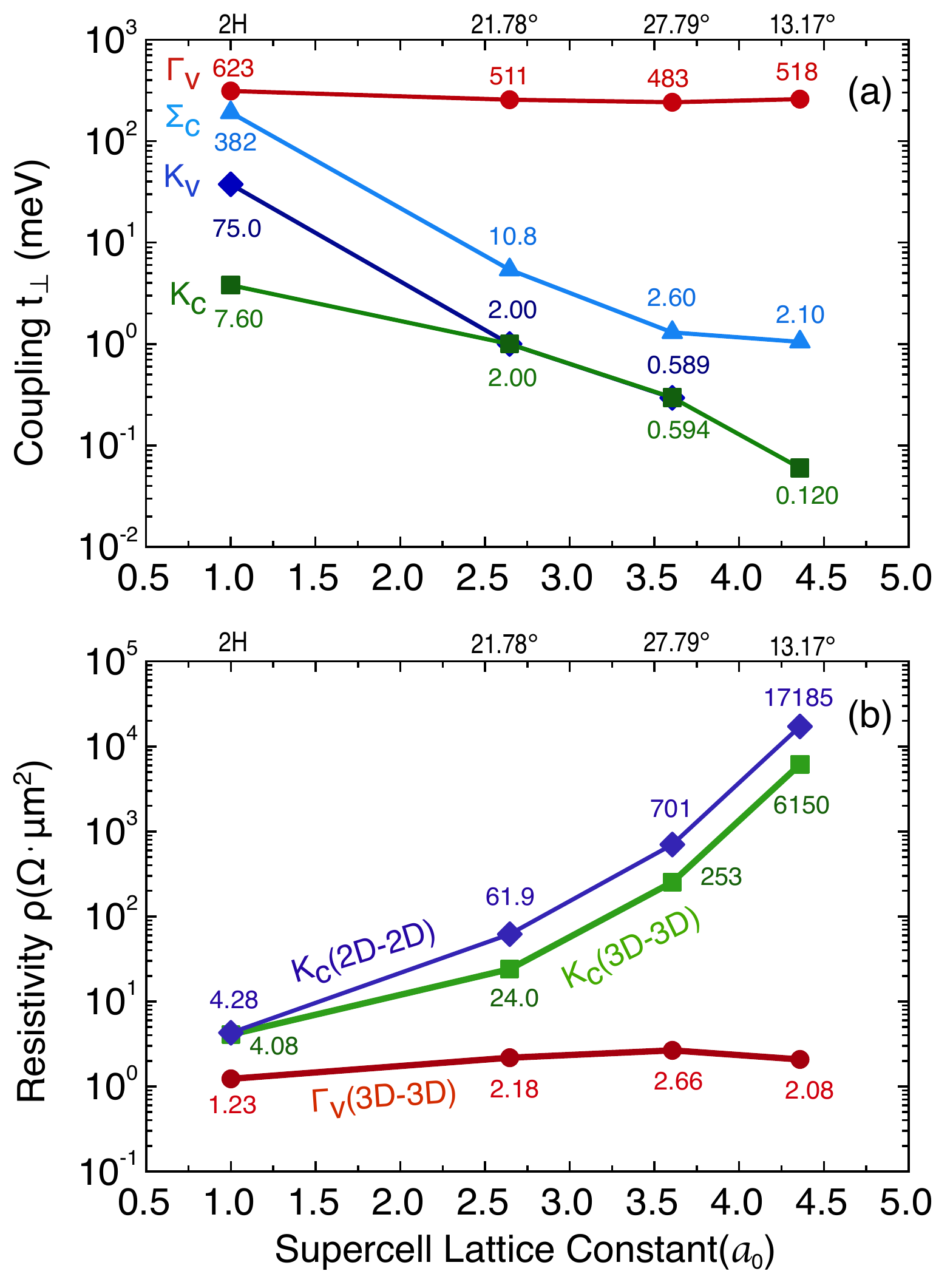}
\caption{
(a) Interlayer coupling $t_\perp$ (meV) of the conduction
band valleys at $K$ and $\Sigma$ labeled $K_c$ and $\Sigma_c$,
and the valence band valleys at $K$ and $\Gamma$ labeled
$K_v$ and $\Gamma_v$. 
(b) Interlayer resistivity ($\Omega \cdot \mu{\rm m}^{2}$)
at the conduction ($K_c$) and valence ($\Gamma_v$)
band edges.
Both the resistivity and the coupling are plotted versus
the commensurate unit cell lattice constant
in units of the unrotated lattice constant $a_0$.
The corresponding angles are shown on the upper horizontal axis.
Numerical values for the data are given next to the
data points.
\label{fig:coupling_rho}
}
\end{figure}

The interlayer couplings are extracted from the energy splittings 
near the band edge as illustrated in Fig. \ref{fig:bands}. 
A semi-log plot of the values versus supercell lattice constant
is shown in Fig. \ref{fig:coupling_rho}(a).
It is clear from Fig. \ref{fig:coupling_rho}(a) that the interlayer
coupling of the holes 
at $\Gamma$ are little affected
by the misorientation angle.
The interlayer couplings of the electron and hole states at
$K$ and $\Sigma$ are exponentially suppressed as a function
of the supercell lattice constant.
This exponential dependence of the band splitting on the  
supercell lattice constant is also found for the 
band splitting in rotated bilayer 
graphene \cite{Avouris_twisted_PRL12}.

Only the conduction $K$ valley and the valence $\Gamma$ valley are considered
for calculating the low-energy electron and hole interlayer resistances, 
since HSE level calculations, which provide more
accurate values for energy levels, 
show that
the conduction band $K$ valley lies approximately 130 meV below the conduction band
$\Sigma$ valley, 
and the valence band $\Gamma$ valley lies 200 meV or more
above the valence band $K$ valley.
\cite{Darshana_JCP14,Louie_Zettl_Twist_MoS2_NComm14,Van_de_Walle_MoS2_PRB14,Darshana_PhD_thesis}.
Once we restrict our attention to the conduction $K$ valley,
which we will denote as $K_c$,
and the valence $\Gamma$ valley, 
which we will denote as $\Gamma_v$,
spin-orbit splitting has little effect on the inter-layer transport,
since the spin splitting of the conduction band at $K$ is 1.5 meV
and the valence band at $\Gamma$ is spin 
degenerate \cite{Burkard_Drummond_Falko_2DMats15}.
Since we are interested in the room temperature conductance,
we ignore the small spin splitting of the conduction band.
To a very good approximation, the low-energy bands within the plane
(perpendicular to the transport direction) are parabolic
and isotropic \cite{Burkard_Drummond_Falko_2DMats15}. 
For the transport calculations,
we treat them as parabolic using two masses, $m_x$ and $m_y$,
such that $\varepsilon(\kv) = \frac{\hbar^2 |\kv|^2}{2m^*}$ with 
$m^* = \sqrt{m_x m_y}$.
The values for the masses from DFT/HSE calculations 
for the holes at $\Gamma$ are $m_x = m_y = 0.62 m_0$
and for the electrons at $K$ are $m_x = 0.47 m_0$ and $m_y = 0.45 m_0$\cite{peelaers2012effects}.

After extracting the interlayer coupling elements $t_{\perp}^\nu$ from the DFT calculations,
we calculate the interlayer conductance of the electron and hole bands using two
different methods described below.
For low energies near a given valley minimum, 
the standard 2D-2D tunneling formula is
\begin{align}
J=\frac { { g }_{ s }g_v q }{ h\mathcal{A} } 
\sum _{\kv} \int  dE \;
&
|t_\perp(\kv)|^2 A_u(\kv;E) A_l(\kv;E) 
\nonumber \\
& 
\cdot \left[ f(E - E_{f,u}) -  f(E - E_{f,l}) \right]
\label{eq:THam1}
\end{align}
where
$A_u(\kv;E)$ is the spectral function of the upper layer,
$A_l(\kv;E)$ is the spectral function of the lower layer,
$t_\perp (\kv)$ is the interlayer coupling determined from
the band splitting,
$f(E-E_f)$ is the Fermi-Dirac factor,
and $E_{f,u(l)}$ is the Fermi level of the upper (lower) layer.
In the prefactor, $\mathcal{A}$ is the area, $g_s$ is the spin degeneracy,
and $g_v$ is the valley degeneracy.
A derivation of this expression from the standard non-equilibrium
Green function expression for the current is given in the Appendix.
The spectral functions are given by 
${A}_{u(l)} = \frac{\gamma}{ \left( E-\varepsilon (\kv) \right)^2 + \frac{\gamma^2}{4}  }
=
\frac{\gamma}{ E_z^2 + \frac{\gamma^2}{4}  }$
where $\gamma$ is the lifetime broadening in each layer
and in the second equality, we define
$E_z \equiv E - \varepsilon(\kv)$. 
Since the interlayer coupling $t_\perp$ is a weak function of $\kv$,
we use its value at the band edge. 
Then, we can perform the sum over the
transverse momenta analytically,
and Eq. (\ref{eq:THam1}) now has the form
\begin{align}
J=\frac { { g }_{ s }g_v q }{ h } & \frac{m^*}{2 \pi \hbar^2}
\int dE_z  |t_\perp|^2 A_u(E_z) A_l(E_z)
\nonumber \\
& \cdot \int_0^\infty d\varepsilon
\left[ f(E_z + \varepsilon - E_{f,u}) -  f(E_z + \varepsilon - E_{f,l}) \right]
\end{align}
For small voltages, the difference in Fermi factors becomes 
$\frac{-\partial f}{\partial \varepsilon} qV$ where $V$ is the applied voltage,
and the integral over $\varepsilon$ gives 
$f(E_z - E_f) qV$ where $E_f$ is the equilibrium Fermi level. 
Therefore, the 2D-2D tunneling formula for the
interlayer conductivity is
\begin{equation}
\sigma_{\rm 2D}=
\frac { g_s g_v q^2 }{ h }
\frac{ m^* }{ 2\pi \hbar^2 }
\int dE_z  
\frac{ |t_\perp|^2 \gamma^2 f(E_z - E_f) }
{ \left[ E_z^2  + \frac{ \gamma^2 }{ 4 }  \right]^2 } .
\label{eq:sigma_2D_2D}
\end{equation}
All calculations of the electronic conductivity $\sigma_{\rm 2D}$ use a value of $\gamma = 12.6$ meV,
and it is estimated from the mobility lifetime using $\mu = e \tau / m^*$ and $\gamma = \hbar / \tau$.
For the mobility, we chose 200 cm$^2$/Vs, which is an average of the best measured value 
for a monolayer of 81 cm$^2$/Vs [\onlinecite{MoS2_1L_natcomm}] and the theoretical value of 
320 cm$^2$/Vs [\onlinecite{KWKim_PRB_TMDC_xport_prop}].

For a given transverse $\kv$, the transmission resulting from this approach is
$T(E;\kv) = \frac{ \left| t_\perp \right|^2  \gamma^2 }{  \left[ { (E-{ \varepsilon(\kv)  }) }^{ 2 }+\frac { { \gamma  }^{ 2 } }{ 4 }  \right]^2  }$
which has a maximum value of 
$16 \left| t_\perp \right|^2 / \gamma^2$.
Since this value must be $\leq 1$, it sets an upper limit on the expression's validity in terms of the 
magnitude of the coupling with respect to the broadening,
$|t_\perp | \leq \gamma / 4 = 3.15$ meV.
For the $K_c$ valley, the unrotated structure does not satisfy this condition, since $t_\perp = 7.6$ meV;
but for all non-zero rotation angles, this condition is satisfied.
For the $\Gamma_v$ valley at all rotation angles, it is not. 

To have an expression that is also valid for the strongly coupled bands,
we create a low-energy, 3D-3D transmission model for each band.
Physically, this corresponds to a system of two semi-infinite stacks
with one stack rotated with respect to the other resulting in the 
rotated interface depicted in Fig. \ref{fig:TBL_MoS2}(a).
For each band, at each transverse $\kv$, this model reduces to that of a 
one dimensional (1D) tight-binding chain as shown in 
Fig. \ref{fig:TBL_MoS2}(b). 
The hopping parameter $t_0$ is given by $t_\perp$ of
the unrotated bilayer in Fig. \ref{fig:bands}(a).
For this model, the `device' consists of the two misoriented layers numbered
0 and 1 in Fig. \ref{fig:TBL_MoS2}(b).
The `device' Hamiltonian for band $\nu$ is given by Eq. (\ref{eq:bilayer_HD}).
The left and right self-energies due to coupling to the 
semi-infinite leads are $\Sigma^R=t_0 e^{ik_za}$.
The Green function is
\begin{equation}
G^R = 
\begin{pmatrix} 
E - \varepsilon_\nu(\kv) - t_0 e^{ik_za}  & -{t}^\nu_{\perp}(\kv) \\ 
-{t}^\nu_{\perp}(\kv) & E - \varepsilon_\nu(\kv)  - t_0 e^{ik_za}  
\end{pmatrix}^{-1} .
\end{equation}
The transmission is calculated from 
$T(E,\kv) = \Gamma_u \Gamma_l |G^R_{0,1}(E,\kv)|^2$
where $\Gamma_l = \Gamma_u = 2 |t_0| \sin(k_za)$.
Using the dispersion relation of the leads,
$E = \ep_\nu(\kv) + 2t_0 \cos(k_z a)$,
this can be analytically evaluated to obtain
$T(E_z) =  \frac{ t_\perp^2 ( 4 t_0^2 - E_z^2 ) }{( t_0^2 + t_1^2 )^2 - t_\perp^2 E_z^2 }$
where $E_z \equiv E - \ep_\nu(\kv)$.
Going through the same steps as for the 2D-2D tunneling formula,
the 3D-3D expression for the conductance is
\begin{equation}
\sigma_{\rm 3D} =
\frac { g_s g_v q^2 }{ h }
\frac{ m^* }{ 2\pi \hbar^2 }
\int_{-2t_0}^{2t_0} dE_z  
\frac{  t_\perp^2 ( 4 t_0^2 - E_z^2 ) f(E_z - E_f) }
{ ( t_0^2 + t_\perp^2 )^2 - t_\perp^2 E_z^2 } .
\label{eq:sigma_3D_3D}
\end{equation}
In all calculations of the interlayer conductance, 
the Fermi level is taken to be $k_BT$ below the conduction
band edge when calculating the electron conductance or $k_BT$
above the valence band edge when calculating the hole conductance,
with T = 300K.
The interlayer resistivity $\rho$ is the inverse of the conductivity
calculated from Eqs. (\ref{eq:sigma_2D_2D}) or (\ref{eq:sigma_3D_3D}). 

{\em Results and Discussion}:
Fig. \ref{fig:coupling_rho}(b) shows 
the interlayer resistivity for electrons
at the conduction band edge at $K$
and the holes at the valence band edge at $\Gamma$.
The interlayer resistivity for holes is only calculated from
the expression for $\sigma_{3D}$ in Eq. (\ref{eq:sigma_3D_3D}),
since the 2D-2D tunneling formula is not valid for the holes
due to the large value of $|t_\perp|$. 
The interlayer resistivity for electrons is calculated from both
expressions, $\sigma_{2D}$ from Eq. (\ref{eq:sigma_2D_2D}) and $\sigma_{3D}$,
and the trends and quantitative values from both expressions match to within
a factor of three over 3 orders of magnitude.
The agreement is not too surprising since the conductivity resulting from
both expressions is proportional to $t_\perp^2$,
and the dependence of the electron and hole interlayer conductivity
follows the dependence of the interlayer coupling shown in 
Fig. \ref{fig:coupling_rho}(a). 

The physics of the interlayer coupling is determined by the 
periodic part of the Bloch function (the orbital composition), 
the phase or envelope $e^{i \kv \cdot \rv}$,
and the interlayer trigonal arrangement of the 3 nearest neighbor Mo atoms
in one layer with respect to a Mo atom in the other layer.
We will first discuss the $K$ valley and then the $\Gamma$ valley.

First, consider the unrotated 2H bilayer.
The very small interlayer coupling of the conduction band is 
due the symmetry
of the conduction band Bloch functions at the $K$ points.
The conduction band edge at $K$ is composed of predominantly Mo $d_{z^2}$ orbitals.
In a minimal basis, the Bloch state at the conduction band edge
of an individual monolayer 
is $\ket{K_c} = \sum_{\Rv_n} \ket{d_{z^2}; \Rv_n} e^{i \Kv \cdot \Rv_n}$
where $\Rv_n$ is the position of each Mo atom.
The conduction band interlayer coupling is proportional to the the interlayer
matrix element 
$\bra{K_c,u}H\ket{K_c,l} = \bra{d_{z^2}^u}H\ket{d_{z^2}^l}  \sum_{n=1}^3 e^{i\Kv\cdot \Rv_n}
\propto \sum_{n=0}^2 e^{in2\pi/3} = 0$
where  $\bra{d_{z^2}^u}H\ket{d_{z^2}^l}$ is the matrix element between
interlayer, nearest neighbor, Mo $d_{z^2}$ orbitals.
Since the interlayer matrix element $\bra{d_{z^2}^u}H\ket{d_{z^2}^l}$ is independent of the 
azimuthal angle, it is pulled outside of the sum, and the sum of the three phase factors exactly cancel.
(For an expanded discussion, see the 
Supplementary Information of [\onlinecite{spin-layer-locking_BL_WSe2_Xu_NatPhys14}].)

In contrast, the valence band state at $K$ is composed of $d_{xy}$ and
$d_{x^2-y^2}$ orbitals. 
The interlayer matrix elements between these
orbitals change sign as a function of the azimuthal angle preventing
the cancellation of the phase factors.
Therefore, at the $K$ valley of the unrotated structure,
even though 
the conduction band orbitals are out-of-plane 
and the valence band orbitals are in-plane,
the interlayer coupling at $K_c$
is an order of magnitude smaller than the interlayer coupling 
at $K_v$, as shown in Fig. \ref{fig:coupling_rho}(a). 

Two mechanisms compete to determine the effect of interlayer 
rotation on the conduction band coupling at $K$. 
When one layer is rotated with respect to the other, the symmetry is broken,
and the exact cancellation of the phases is destroyed. 
This effect would cause the matrix element to increase.
However, now the unit cell size has increased to one of the supercells
shown in Fig. \ref{fig:TBL_MoS2}, and the interlayer matrix elements between all 
of the $d_{z^2}$ orbitals in the supercell and their associated phase factors must be added. 
At $K$, the phase is changing sign approximately every lattice constant,
so that as the wavefunction of the top layer is rotated with respect to that of the
bottom layer, and the phases are summed over the large supercell, 
the matrix element is suppressed by phase cancellation.
These two competing effects cause the initial slower decrease in the coupling
of the conduction band at $K$ compared to the coupling of the valence band at $K$
as shown in Fig. \ref{fig:coupling_rho}(a).

The effect of misorientation on 
the the interlayer resistivity of the electrons at $K$  
is similar to the effect of misorientation on the 
interlayer resistivity of electrons and holes in 
bilayer graphene \cite{bistritzer2010transport,Avouris_twisted_PRL12,Habib_Graphite_APL13}.
The electron resistivity increases exponentially with the size of the 
supercell lattice constant, although the increase in MoS$_2$ is orders
of magnitude less than the increase in bilayer graphene
(compare Fig. 1d of [\onlinecite{Avouris_twisted_PRL12}] or 
Fig. 4 of [\onlinecite{Habib_Graphite_APL13}] with Fig. \ref{fig:coupling_rho}(b)).

The valence band edge at $\Gamma$ is composed of 28\% S $p_z$ orbitals
and 67\% Mo $d_{z^2}$ orbitals\cite{Darshana_JCP14}. 
These out-of-plane orbitals, especially
the $p_z$ orbitals on the surface S atoms, strongly couple between layers.
Furthermore, the interlayer matrix elements are independent of the 
azimuthal angle, and at $\Gamma$, all of the phase factors are 1, 
so the matrix elements add, and the interlayer coupling is large
as shown in Fig. \ref{fig:coupling_rho}(a).

When one layer is rotated with respect to the other,
no phase cancellation can occur, since the $\Gamma$ wavefunctions have no phase.
Thus, the holes at $\Gamma$ are minimally affected by layer rotation.
The only effect on the hole coupling is through the slight increase
in the interlayer separation causing a slight decrease
in the interlayer coupling as shown in Fig. \ref{fig:coupling_rho}(a).
Furthermore, the interlayer coupling of the holes monotonically 
decreases with angle rather than with supercell size, 
following the monotonic increase of the interlayer distance 
(see Table S1),
which further indicates that different physics
govern the effect of misorientation on the electron and hole interlayer coupling.

To gain perspective into what the resistivity values mean for a
device application, we consider the target resistivity value
of 2.5 $\Omega \mu$m$^2$ for the emitter contact resistance 
required to achieve THz cutoff frequency in a heterostructure bipolar
transistor (HBT) \cite{Rodwell_ProcIEEE08}. 
The interlayer resistivity of the holes is approximately equal to or below that value for all 
angles.
For all non-zero rotation angles considered,
the interlayer resistivity of the electrons is one or more orders of magnitude too high.
This suggests design optimization of a
heterostructure bipolar transistor (HBT) using stacked TMDs.
A pnp HBT will be insensitive to misalignment of the layers. 
Furthermore, rotating the emitter layer with respect to the base layer in a pnp HBT
will increase the emitter injection efficiency by one or more orders of magnitude, 
since the transmission of electrons injected from the base
will be suppressed while the 
transmission of holes injected from the emitter will be unaffected.

{\em Conclusion}:
The electron interlayer coupling of the conduction band at $K$ is weak
(7.6 meV),
and it decreases by a factor of 63 as the supercell lattice constant
increases by a factor of 4 corresponding to a $13^\circ$ rotation.
The hole coupling is large (632 meV) and remains large decreasing by a factor of 1.3
at a rotation angle of $27.8^\circ$
The corresponding electron interlayer resistivity
increases from 4 $\Omega \mu {\rm m}^2$ to $10^4$ $\Omega \mu {\rm m}^2$.
The hole resistivity remains near $2$ 
$\Omega \mu {\rm m}^2$ for all rotation angles.
Interlayer rotation between an n-type and p-type region will suppress the electron
current, which is desirable in the base-emitter junction of a pnp HBT.

\noindent

{\em Acknowledgement}: This work was supported by FAME,
one of six centers of STARnet, a Semiconductor Research Corporation
program sponsored by MARCO and DARPA. This work used the Extreme 
Science and Engineering Discovery Environment (XSEDE), which is supported 
by National Science Foundation grant number ACI-1053575.

\appendix*

\section{Ab-initio simulation details and derivation of Eq.(2)}


%
Electronic structure calculations of bilayer MoS$_2$ are 
carried out using density functional theory (DFT) with a 
projector augmented wave method and the Perdew-Burke-Ernzerhof (PBE) 
type generalized gradient approximation as implemented in the 
Vienna Ab initio Simulation Package 
(VASP) \cite{VASP:I:PRB:1993,VASP2,VASP:II:PRB:1996,perdew:1996:PBE,Ernzerhof_JChem99}.
A semi-empirical Grimme-D2 correction to the Kohn-Sham energies 
is used to model the van der Waals (vdW) interactions \cite{Grimme_DFT_D2}. 
Spin-orbit coupling is not included, since it has little effect on the interlayer coupling parameter, which
is determined by orbital overlap.
The plane wave basis energy cutoff is 400 eV.
The global break condition for the electronic SC-loop is below ${10}^{-6}$ eV.
The Monkhorst-Pack scheme is used for the integration over the 
Brillouin zone with a 
$\Gamma$ centered 
k-mesh of $12\times12\times1$ for the unrotated thin films. 
For rotated bilayers, k-mesh are accordingly revised to $3\times3\times1$ for ${13.17}^{ \circ }$, 
$6\times6\times1$ for ${21.78}^{ \circ  }$, $4\times4\times1$ for ${27.79}^{ \circ }$, 
since they have different Brillouin zones.
The k-space integration was carried out with a Gaussian smearing 
width of 0.02 eV for all calculations.
All unit cells were built with 20 $\AA$ separation between replicas in the perpendicular direction to achieve negligible interaction.

The default optimization methods did not efficiently 
determine the bilayer separation because the van der Waals 
interaction energies are very small.
In order to accurately determine the bilayer separation of each 
system, several specific layer separations were used 
to optimize the structures until all or the interatomic forces are below 0.01 eV/$\AA$ 
as described in the Supplementary Information of [\onlinecite{Hone_Heinz_Tailoring_Twist_MoS2_NL14}]. 
The optimized structure with lower total energy was chosen for structure of each rotated angle.  
%
%
The interlayer distances of the relaxed structures 
are shown in Table \ref{tab:d_and_Eg}.
%
\begin{table}[!hbp]
\renewcommand\thetable{S1}
\centering
\begin{tabular}{ccccc}
\hline
Angle $\theta$ & $a/{a}_{0}$  &d ({\AA})  & $E_{\Gamma-K}$(eV)   & $E_{K-K}$(eV) \\
\hline\hline
2H(0)      & 1.0               &6.2568     &  1.3650    & 1.7348   \\
\hline
${13.17}^{ \circ  } $     & $\sqrt{19}$  & 6.5142    &  1.5279     & 1.7683   \\
\hline
${21.78}^{ \circ  } $     & $\sqrt{7}$    & 6.5287    &  1.5339     & 1.7669    \\
\hline
${27.79}^{ \circ  } $      & $\sqrt{13}$  & 6.5853    &  1.5617     &  1.7698   \\
\hline
\end{tabular}
\caption{
Interlayer distance (d), supercell lattice constant $a$, 
indirect energy gap $E_{\Gamma-K}$, 
and direct gap $E_{K-K}$ as a function of rotation angle.
}
\label{tab:d_and_Eg}
\end{table}

The standard 2D-2D tunneling formula can be obtained following
the derivation leading to the 
current expression of Meir and Wingreen \cite{Meir_Wingreen}
\begin{align}
J=\frac { { g }_{ s }g_v q }{ hA } 
\sum _{\vec {k}} \int { dE } 
{\rm tr} 
\{
\Gamma_u (E;\kv) 
&
[ f(E - \mu_u) A_l(E;\kv)  
\nonumber \\
&
+ i G_l^<(E;\kv)  ]
\} .
\label{eq:M_and_W}
\end{align}
For the system shown in Fig. 1(a) of the main text,
$A_l$ and $G_l^<$ are the spectral function
and less-than correlation function of the lower layer, and
$\Gamma_u(E;\kv) = t_\perp^2(\kv) A_u(\kv;E)$, where
$A_u(\kv;E)$ is the spectral function of the upper layer.
In a tunneling Hamiltonian approach, the two layers are assumed to be weakly
coupled, so that each layer can be approximated as equilibrated with its
own Fermi level.
Then, $G_l^<(\kv;E) = i f(E-\mu_l) A_l(\kv;E)$, and Eq. (\ref{eq:M_and_W}) becomes
\begin{align}
J=\frac { { g }_{ s }g_v q }{ hA } 
\sum _{\kv} \int  dE \;
&
|t_\perp(\kv)|^2 A_u(\kv;E) A_l(E;\kv) 
\nonumber \\
& 
\cdot \left[ f(E - \mu_u) -  f(E - \mu_l) \right] .
\label{eq:THam1_app}
\end{align}




%

\end{document}